\documentclass{emulateapj}

\newcommand{\kms}{\ensuremath{\,\mbox{km}\,\mbox{s}^{-1}}}

\newcommand{\Msun}{$M_{\odot}$}
\newcommand{\HI}{H{\sc i}\ }

\begin{document}
\title{The Star Formation Threshold in NGC 6822}
\author{W.J.G.~de~Blok}
\affil{Research School of Astronomy \& Astrophysics, ANU\\
Mount Stromlo Observatory, Cotter Road, Weston Creek, ACT 2611, Australia}
\email{edeblok@mso.anu.edu.au}
\and
\author{F. Walter}
\affil{Max Planck Institute f\"ur Astronomy\\
K\"onigstuhl 17, 69117 Heidelberg, Germany}
\email{walter@mpia.de}
 
\begin{abstract}
We investigate the star formation threshold in NGC 6822, a nearby
Local Group dwarf galaxy, on sub-kpc scales using high-resolution,
wide-field, deep H{\sc i}, H$\alpha$ and optical data.  In a study of
the \HI velocity profiles we identify a cool and warm neutral
component in the Interstellar Medium of NGC 6822. We show that the
velocity dispersion of the cool component ($\sim 4$ \kms) when used
with a Toomre-$Q$ criterion gives an optimal description of ongoing
star formation in NGC 6822, superior to that using the more
conventional dispersion value of $6$ \kms.  However, a simple constant
surface density criterion for star formation gives an equally superior
description.  We also investigate the two-dimensional distribution of
$Q$ and the star formation threshold and find that these results also
hold locally.  The range in gas density in NGC 6822 is much larger
than the range in critical density, and we argue that the conditions
for star formation in NGC 6822 are fully driven by this density
criterion. Star formation is local, and in NGC 6822 global rotational
or shear parameters are apparently not important.
\end{abstract}


\section{Introduction\label{THRESHOLD}}

In many galaxies, star formation as traced through H$\alpha$ emission
is usually only found in the inner parts, whereas the \HI disk is seen
to extend to much larger radii.  The presence of lower column density
gas at large radii, coupled with the absence of obvious star formation
there has led to the concept of a star formation threshold: below a
certain critical density (which can be constant or varying with
radius, depending on the assumptions made) gas is unable to turn into
stars.  The idea of a star formation threshold has been pursued both
observationally and theoretically.  For example, \citet{skillman87}
found that a constant \HI column density value of $\sim 10^{21}$
cm$^{-2}$ separated higher column density star forming regions from
lower density inert regions.  \citet{kennicutt89} and more recently
\citet{martin01} developed the idea of a star formation threshold
originating in the Toomre $Q$ parameter describing gravitational
instabilities in a gaseous, rotating disk (see also
\citealt{quirk72}).  This description is based on comparisons of
azimuthally averaged CO/\HI and H$\alpha$ profiles, and thus provides
a \emph{global} picture of the threshold.  Note that in some galaxies
isolated H{\sc ii} regions are found in regions with gas densities
below the (azimuthally averaged) threshold value
(e.g.\ \citealt{ferguson98}): these may occur at positions where the
\emph{local} conditions are favourable to star formation.  The recent
discovery of recent star formation in the outer disks of M83
\citep{thilker05} and NGC 4625 \citep{depaz05} additionally suggests
that H$\alpha$ may not be a complete tracer of ongoing star fomation
in disk galaxies. Other evidence for extended star formation in
irregular dwarf galaxies by tracing blue ``clumps'' is presented in
\citet{parodi}.

In this paper we investigate the star formation threshold in NGC 6822.
In a companion paper [de Blok \& Walter 2005 (Paper I)] we have
presented high-resolution H{\sc i}, H$\alpha$ and optical datasets
that give for the first time a detailed picture of the distribution of
gas, recent star formation and the resolved stellar population in this
galaxy.

We refer to Paper I for a description and presentation of these data.
In summary, the \HI data set consists of an 8-pointing mosaic obtained
with the Australia Telescope Compact Array.  The data have a spatial
resolution of $42'' \times 12''$ and a velocity resolution of 1.6 \kms
and show the presence of an extended \HI disk. The optical data
consists of $B$, $R_C$, and $I$ observations obtained with Suprime-Cam
on the Subaru telescope. These resolve the stellar population in NGC
6822.  By isolating stars with $(B-R)<0.75$ and $18 \la m_B \la 26$
(no extinction corrections applied), we show the presence of a young,
blue population spread throughout the entire extended \HI disk in a
highly structured fashion.  This population consists of main sequence
stars with luminosities $-7 \la B \la +1$, corresponding to O to A
stars with main-sequence life-times from $\sim 1$ to $\sim 100$ Myr.
Our wide-field H$\alpha$ data (obtained with the Widefield Camera on
the Isaac Newton Telescope) show the presence of several hitherto
unknown H{\sc ii} regions in the outer HI disk.

Here we bring these data sets together in a study of the star
formation threshold. In many ways NGC 6822 is an ideal galaxy for
testing the various explanations and incarnations of the star
formation threshold. It is a nearby Local Group dwarf galaxy
unaffected by the presence of a bulge or spiral arms that could
complicate the interpretation of its dynamics.  Its small distance of
only 490 kpc enables us to study the Interstellar Medium (ISM) and the
stellar population on scales of $\sim 100$ pc or less, i.e., the scale
of Giant Molecular Clouds.

In the next section we discuss various observational and theoretical
interpretations of the threshold. Section~3 deals with the
decomposition of the \HI velocity profiles and the identification of a
cool and warm component of the neutral ISM in NGC 6822. Section 4
discusses the radial as well as the two-dimensional distribution of
the star formation threshold. We summarise our results in Sect.~5.

\section{Interpretations of the star formation threshold\label{SEC2}}

The most popular explanation of the star formation threshold is based
on the Toomre $Q$ parameter.  This parameter is defined as
\begin{equation}
Q(r) = {{c_s \kappa}\over{\pi G \Sigma_g}},
\end{equation}
where $c_s$ is the sound speed in the gas, $G$ the gravitational
constant, $\Sigma_g$ the surface density of the gas and $\kappa$ is
the epicyclic frequency, defined as $\kappa^2 = 2
(V/R)^2[1+(R/V)(dV/dR)]$. Itself, $\kappa$ is a function of the
rotation curve $V(R)$.  In this picture, a disk will be unstable
against radial perturbations if $Q<1$, and one can define a critical
density for star formation
\begin{equation}
\Sigma_c = \alpha {{c_s \kappa}\over{\pi G}}.
\end{equation}

The constant $\alpha$ is used to reconcile the model with the
observations.  In most analyses, including this one, the
one-dimensional velocity dispersion $\sigma$ is used instead of
$c_s$. \citet{joop} makes the explicit correction $c_s = \sigma
\sqrt{\gamma}$, where $\gamma = 5/3$ is the ratio of specific heats
for an adiabatic mono-atomic gas. We prefer to use $\sigma$, keeping
in mind that the sound speed correction is taken up into the value of
$\alpha$.

In essence, Eqs.~(1) and (2) describe the ability of perturbations to
rotate around their centre, and thus their stability against
collapse. \citet{kennicutt89} applied this prescription to a number of
spiral galaxies and compared the radial distributions of the ratio of
the critical density $\Sigma_c$ and the neutral gas density $\Sigma_g$
with the distribution of H$\alpha$ in 15 spiral galaxies. He found
that for a value of $\alpha = 0.63$ he could correctly describe the
steep drop-off in star formation at the H$\alpha$ truncation radius
for most galaxies. This work was later extended and confirmed by
\citet{martin01} who studied a sample of 32 nearby spiral
galaxies. Exceptions do exist though: \citet{martin01} point out that
a few nearby spirals with active and organised star formation
(e.g.\ NGC 2403 and M33) have sub-critical densities throughout their
disks.

In the context of a critical threshold, star formation is often
thought to be self-regulatory and striving for a $Q\sim 1$
situation. Regions where $Q<1$ will form stars, diminishing the local
gas density, and increasing the local velocity dispersion due to
mechanical stirring and true heating.  All this will eventually quench
star formation. A new build-up of gas, e.g.\ due to nearby star
formation, can then restart the whole process again.

The critical threshold is also used to explain the low star formation
rate found in LSB galaxies. \citet{vanderhulst93} found that in these
galaxies the gas surface densities were below the critical density
over most of their disks. Local density enhancements were thought to
provide the necessary instability for stochastic star formation.

An alternative description of the threshold is given by
\citet*{heb98}.  They analysed a sample of dwarf and irregular
galaxies and needed a value of $\alpha$ which was a factor $\sim 2$
lower than the \citet{kennicutt89} value to explain their results in
the context of a $Q$ parameter (though part of this difference is due
to the different velocity dispersion value they used).  They argue
that for their galaxies the local shear rate gives a much better
description of the ability of perturbations to collapse. This shear
threshold is defined in a similar way as the threshold in Eq.~(2),
with $\kappa$ replaced by $\pi A$, where $A = -0.5\, R\,
(d\Omega/dR)$, one of the Oort constants, where $\Omega$ is the
angular speed.  

Much effort has also been put into finding a physical
basis for the star formation threshold. The most promising one seems
to involve the phase structure of the ISM, as pointed out by
\citet{elmegreen94}, and emphasized by
\citet{jeroen97,heb98,hunter01,billet02} and \citet{elmegreen02}.
Recently, several of the observational and theoretical strands were
drawn together in a paper by \citet{joop}.  Its major points that are
relevant for the current analysis can be summarised as follows.

The main conclusion from \citet{joop} is that star formation can only
happen if the ISM contains a cool phase (see also \citealt{ned}).  The
transition to a cool phase is associated with a steep drop in the
thermal velocity dispersion, which causes instabilities on a wide
range of scales, and is the physical cause for the drop in $Q$ at the
critical radius in the models.  This implies that in determining $Q$
the relevant parameter is the thermal velocity dispersion of the cool
phase of ISM.  The dispersion that is usually used, as derived from
straight-forward second-moment maps, is predominantly caused by the
warm neutral ISM, and will therefore usually be higher than that of
the cool component.  The simple second-moment velocity dispersion
values are therefore not necessarily applicable to a star formation
threshold analysis.

In local hydrostatic equilibrium with only a warm phase present, there
is a direct relationship between density, surface density and
pressure. For fixed temperature, gas fraction, metallicity and UV
field the formation of a cool phase depends only on density, and
therefore surface density. Flaring of the disk is, in these models,
only expected to occur beyond the critical radius, where the
temperature increases steeply; for the purposes of this discussion the
scale height of the disk is thus constant within the critical radius.
In the models, the drop in velocity dispersion thus occurs at a fixed
critical surface density.  This surface density agrees with the values
derived observationally.  Rotation (i.e.\ Coriolis force and shear)
\emph{cannot} prevent a cool phase from forming within the critical
radius: the rotation curve and epicyclic frequency change smoothly
across this radius. As the threshold therefore only depends on surface
density, it must be a local phenomenon.  Any peak that exceeds the
critical surface density has the potential to form stars, regardless
of its position.  As \citet{joop} notes, the model predicts its own
demise: once stars form, feedback in a complex multi-phase ISM is not
well described by a simple model. One should thus not necessarily
expect the critical density recipe to work in regions of active
starformation.  In the models, the critical hydrogen surface density
for star formation is $\log(N_H) \simeq 20.75$ ($\sim$ 4.5 $M_{\odot}$
pc$^{-2}$), but depends slightly on metallicity, ambient radiation
field, and gas fraction.

Lastly, the reason why the conventional interpretation of the
threshold with its (too high) constant velocity dispersion does yield
correct values for the critical radii, is shown in Fig.~3 of
\citet{joop}.  The change in $Q$ with radius is shown there, both for
the choice of velocity dispersion motivated by the model, as well as
the conventional assumption of a constant velocity dispersion.  The
latter choice gives a $Q$ that increases gradually with radius, and at
the critical radius intersects the steeply dropping model $Q$ value.
As the drop in this ``true'' $Q$ is so steep and sudden, the
conventional $Q$ curve will always intersect it at the critical
radius, independent of the choice of observed velocity dispersion.  It
will furthermore do this at $Q=1/\alpha$. Choosing the appropriate
value for the velocity dispersion (namely that of the cool phase), and
making the correction for the sound speed (see Sec.~\ref{SEC2}) makes
the intersection happen at $Q=1$.  For example, the
\citet{kennicutt89} value of $\alpha=0.63$ was derived assuming a
dispersion of 6 \kms. Correcting this for the sound speed, yields
$\alpha \simeq 0.5$. Using a lower velocity dispersion of $3-4$ \kms,
as appropriate for the cool phase, leads to $\alpha \simeq 1$.

In summary, an increase in density in the warm phase leads to an onset
of a cool phase, accompanied by a steep drop in velocity dispersion,
resulting in a lower value of $Q$. This occurs at a fixed surface
density, and does not depend on rotation. The star formation threshold
is a local phenomenon.

\section{HI profiles and velocity dispersion\label{SEC:GAUSPROF}}

The integrated \HI map and the velocity field of NGC 6822 are presented
in Paper I (see also Weldrake et al.~2003). They show an extended \HI
disk characterized by the presence of the NW companion --- a dwarf galaxy
interacting with the main body of NGC 6822 --- and a super-giant shell,
presumably caused by the effects of star formation, dominating the SE
part of the disk. The velocity field shows that the galaxy is in
regular solid-body rotation.

From the \HI data presented in Paper I we can also derive a
second-moment map (where the second moment is defined as
$\sqrt{\langle V^2 \rangle - \langle V \rangle^2}$). This map is shown
in Fig.~\ref{fig:mom2}.  It should be compared with the
integrated HI map and velocity field as well as the maps of the
stellar distribution presented in Paper I.  The highest second-moment
values ($ > 8$ \kms) are found in the region roughly corresponding
with the optically prominent part of the disk, whereas the NW companion,
the large hole and the SE tidal tail are characterised by values $< 8$
\kms.  The median second-moment value in the \HI disk is 6.7\ \kms.
For simple, single-component Gaussians, these second-moment values can
be directly interpreted as physical dispersion values.  However, for
more complicated profiles, second-moment values, even though often
still called ``dispersions'', should be treated with care.  For
non-Gaussian or multi-component profiles (which are ubiquitous in NGC
6822 as we will show below) the interpretation of second-moment values
as dispersions is not always unambiguous.  Furthermore, in a
multi-phase medium with star formation the measured velocity
dispersion is the sum of the dispersions of the cool and warm phase
with additional input from star formation and turbulence. As
the relevant dispersion value to use in a star formation threshold
analysis is that of the cool phase of the ISM, using the second-moment
values will likely result in significant over-estimates.  A simple
second-moment analysis is  not sufficient for the current
investigation, and a more precise analysis of the HI velocity profiles
is needed.

\subsection{High-velocity gas\label{HVC}}

In order to study the threshold for star formation by means of the \HI
velocity profiles it is essential to concentrate only on those parts
of the galaxy where the analysis is valid.  Regions where the ISM is
heavily disturbed are less suited, as the velocity dispersion is less
likely to reflect the intrinsic state of the ISM, but rather bulk
motions of gas along the line of sight.  Disturbed gas can usually be
identified by velocities that significantly deviate from the local
rotation velocity.

To locate this high-velocity (HV) gas in NGC 6822 we have used the
data cube presented in Paper I, and isolated all emision brighter than
3 times the RMS noise level and more than 18 km s$^{-1}$ (or
3$\sigma$, assuming $\sigma= 6$ km s$^{-1}$) away from the local rotation
velocity\footnote{Note that here we have used the conventional
  dispersion of $6$ \kms, rather than that of any cool or warm ISM
  component.  The reason is that here we are interested in the
  properties and behaviour of the ISM as a whole, not just its cool
  component.}.

The top-left panel in Fig.~\ref{fig:HVC} shows the location of this HV
gas overlaid on the normal \HI surface density map of NGC 6822.  We
distinguish between HV gas with positive and negative deviation from
the rotation velocity, but Fig.~\ref{fig:HVC} shows that the majority
of the HV gas has a positive deviation.

The top-centre and top-right panels in Fig.~\ref{fig:HVC} compare the
position of the HV gas with the distributions of the H$\alpha$
emission and the young, blue population (as described in Sect.~1 and
Paper I), respectively. It is obvious that most of the HV gas is
associated with H$\alpha$, though not necessarily with the strongest
emission. This contrasts with the distribution of the young, blue
stars, which seems to be anti-correlated with the HV emission.  The
H$\alpha$ emission has a clear minimum at the positions of the two
most prominent concentrations of blue stars at the tips of the central
bar-like structure. These correlations could indicate a rapid infall
of any HVC gas back into the disk, or may be an indication for
ionization of the gas.

The most prominent HV feature is a partially ring-like structure found
in the southern part of the disk.  Its signature is consistent with
that of chimney due to star formation.  The proximity to the so-called
Blue Plume (indicated in the top-right panel in Fig.~\ref{fig:HVC}),
one of the most active regions of star formation in NGC 6822, make it
likely that the combined effects of star formation in this area have
caused this feature.  The similarity in size between the Plume and the
chimney supports this idea.  If this association is correct, then the
chimney is not caused by a single star cluster, but by a large
conglomerate of young stars $\sim 0.5$ kpc in diameter.

The  bottom  panel  shows  a  number  of cross-cuts  in  the  form  of
position-velocity diagrams.  As we step along the slices from north to
south we  first see the  appearance of a  single component of  HV gas,
which  then splits up  in what  appear to  be two  walls of  a chimney
(slice 5),  which then  merge again into  a single  component.  Beyond
this, there is an extended tail  of HV gas which in projection extends
beyond the \HI disk, but which in reality is likely to have been removed
vertically from the chimney.

If the latter is the case, and given that the deviation with respect
to the rotation velocity is positive, the assumption that the gas is
still positioned more or less over the chimney, then leads to the
conclusion that the HV gas is situated on the far side of the
disk. This would mean that the southern part of the disk of NGC 6822
is the part closest to us.  We can use the average inclination and PA
values as given in \citet{weldrake03} ($i\simeq 60^{\circ}$ and PA
$\simeq 135^{\circ}$) to show that the HV gas plume must then be $\sim
170$ pc above the disk. Comparing with the value for the \HI scale
height of 0.28 kpc derived in \citet{deblokwalter00}, shows that this
gas is still close to the disk, and will most likely fall back.

\subsection{Profile Decompositions}

To extract the \HI velocity profiles we Hanning-smoothed the cube to
reduce the impact of Galactic residuals, and regridded to a pixel size
of $48''$, thus ensuring that every spatial pixel contains an
independent profile.  To isolate the positions containing sufficient
signal for this analysis, we restrict ourselves to profiles whose peak
flux is four or more times the RMS noise per channel.

Profiles were fitted with both a one-component and a two-component
Gaussian model.  The majority of the profiles turned out to be
non-Gaussian.  A comparison between the goodness-of-fit parameters of
both sets of fits showed that of the 732 profiles, 241 could be fitted
equally well with either one or two components (that is to say, their
reduced $\chi^2$-values agreed to within 10 percent). For 438 profiles
the two-component fit produced significantly lower reduced $\chi^2$
values than the single component fit (i.e.\ a difference larger than
10 percent). For 53 profiles a one-component fit was preferred. 
The formal error in the dispersion for the two-component fits is 
$\sim 0.6$ \kms, much less than the dispersions measured.

Figure \ref{fig:chi2comp} compares the quality of the one- and
two-component fits. The dots indicate the positions of the profiles.
The profiles where the two-component fit is preferred are spread
across most of the galaxy. Profiles where a one-component fit performs
better are only found in the SE tidal arm. This is also the region
where residual Galactic foreground emission affects the velocity
profiles significantly (cf.\ Fig.\ 2 in Paper I), and this, rather
than any physical explanation, is the cause of the better performance
of the one-component fits. In the following we will therefore stay
clear of a physical interpretation of the profiles in the tidal arm.

The profiles where one- and two-component fits perform equally well
tend to be found towards the outer parts of the disk, notably the
northern edge of the NW companion and the eastern rim of the
supergiant shell. However, there is no clear correlation between the
presence of e.g.\ star forming regions and number of components. The
only obvious conclusion one might draw is that the two-component fits
tend to be preferred in regions with a higher blue stellar surface
density (compare Fig.~\ref{fig:chi2comp} with the top-right panel of
Fig.~\ref{fig:HVC}).

Figure \ref{fig:prof_panel} shows a representative subset of profile
shapes. Most of them are non-Gaussian and seem to consist of a narrow
component superimposed on a broader one (e.g.\ the profiles in the
bottom-right of Fig.~\ref{fig:prof_panel}), while others are
asymmetric (top-left) or even double-peaked (bottom-left).

Figure \ref{fig:prof_example} shows an example of a profile where a
two-component model is clearly preferred. Inspection of the individual
fit-components shows that one is characterized by a large velocity
dispersion $\sigma \sim 12$ km s$^{-1}$, while the other has a narrow
dispersion $\sigma \sim 5$ km s$^{-1}$.  This is a general feature of
the profiles: for the majority a decomposition using a two-component
Gaussian model results in a broad and a narrow component.

Figure \ref{fig:prof_dispersion} shows histograms of the velocity
dispersions of both components, superimposed on the distribution
derived from a one-component Gaussian fit.  The latter is very close
to the distribution of dispersions one would derive from the
second-moment map (Fig.~\ref{fig:mom2}).  The narrow component in
Fig.~\ref{fig:prof_dispersion} has a mean velocity dispersion value of
4.4 km s$^{-1}$ with an RMS of 1.8 km s$^{-1}$. The broad component
has a mean of 8.2 km s$^{-1}$ with an RMS of 2.4 km s$^{-1}$.  There
is some overlap between the two components, this is mostly due to
profiles that are double peaked (such as the ones in the centre-left
of Fig.~\ref{fig:prof_panel}). Here the fit does not capture narrow
and broad components but rather traces the bulk motion of gas clouds
along the line of sight due to e.g.\ effects of star formation.

The small size of the beam with respect to that of the galaxy (48$''$
equals 110 pc at the NGC 6822 distance) and the shallow rotation curve
rule out that we are looking at lines that are rotationally
broadened. With an inclination of $\sim 60^{\circ}$ the maximum
projected beam size in the plane of NGC 6822 is 220 pc, small enough
to rule out that we are looking at kinematically uncorrelated
components within the beam.  Due to the peak flux cut-off used in
making the fits, the total \HI mass recovered by the fits is smaller
than the total \HI mass of NGC 6822. The two component fits retrieve
71 per cent of the total mass, or $9.5 \cdot 10^7$ \Msun. Of the
recovered mass, 29 per cent (or $2.8 \cdot 10^7$ \Msun) forms the
narrow dispersion component, while 71 per cent ($6.7 \cdot 10^7$
\Msun) is found in the broad dispersion component.

Most of the overlap between the histograms in
Fig.~\ref{fig:prof_dispersion} is due to double-peaked profiles.
Based on Fig.~\ref{fig:prof_dispersion} we thus introduce an
alternative division between the narrow and broad component by
defining as ``narrow'' all those profiles with $\sigma \le 6$ km
s$^{-1}$ and as ``broad'' those with $\sigma > 6$ km s$^{-1}$.  Using
this alternative definition 19 per cent of the retrieved \HI (or $1.8
\cdot 10^7$ \Msun) forms the narrow dispersion component, while 81 per
cent ($7.7 \cdot 10^7$ \Msun) is found in the broad dispersion
component.  These ratios are very similar to the ratios found in
previous studies of broad and narrow components in dwarf galaxies
(e.g., \citealt{younglo96, younglo97}).

\subsection{Broad and narrow components}

Figure \ref{fig:warmcool} shows the distribution of the narrow and
broad components, using the 6 km s$^{-1}$ definition. The lack of
emission in the SE corner (corresponding to the location of the tidal
tail) is due to the presence of residual Galactic foreground emission
making the interpretation of the fits there problematic
(cf.\ Fig.~\ref{fig:chi2comp}).  In the following we will concentrate
on the emission in the other, unaffected parts of the area.  The broad
component seems more widely, and more continuously distributed.  A
slight anti-correlation between the components seems present: most
(though not all) of the high-column density narrow dispersion emission
is found at positions of lower broad-dispersion intensity, and vice
versa.

Figure \ref{fig:warmcool} should be interpreted with care though, and
compared with Fig.~\ref{fig:HVC}.  In most of the disk the
decomposition in two components reflects the local state of the
neutral gas component, but as Fig.~\ref{fig:HVC} shows, in regions in
the central parts of the disk where double-peaked profiles are found,
the interpretation is less straight-forward. For example, the
prominent peak in the low-dispersion component around $(\alpha,
\delta) =$ $19^h44^m50, -14^{\circ}55'40''$ occurs in a region where a
significant high-velocity gas component is present (see
Sect.~\ref{HVC}).

In all of this it should be kept in mind that the resolution of the data
and the rebinning to 48$''$ may lead to an underestimate of the importance
of the narrow dispersion component, as does the complex velocity structure
in the inner parts.

The narrow dispersion component has a patchy distribution but is
generally found throughout the star forming disk.
Fig.~\ref{fig:warmcool} also compares the distribution of the narrow
dispersion component with that of the H$\alpha$ emission and the blue
stars.  Except for the aforementioned prominent peak where
interpretation is made difficult by the presence of high-velocity gas,
every other low-dispersion peak has a compact H$\alpha$ region in its
direct environment (though the reverse is not always true). This even
holds for the low surface brightness H$\alpha$ regions in the NW
companion and on the SE rim of the giant hole.  The narrow dispersion
component also seems to avoid regions of high densities of young
stars.

In the following we will associate the narrow dispersion component we
have identified in NGC 6822 with the cool phase of the neutral ISM. We
will adopt an average value of 4~\kms\ as the velocity dispersion of
the cool component in our analysis of the star formation threshold
(see Fig.~\ref{fig:prof_dispersion}).

\section{The Critical Density}

\subsection{Radial Distributions of Critical Density}

Figure \ref{fig:thresh_rad} compares the radial distribution of the
neutral gas with that of the critical density as defined in
\citet{kennicutt89} and the shear critical density investigated in
\citet{heb98}. For both we have used the $12''$ rotation curve and \HI
surface density profile presented in \citet{weldrake03}. The gas
density was derived from the inclination-corrected \HI surface density
profile by multiplying the latter with 1.4 to correct for the presence
of helium and metals. We show the critical densities for a dispersion
value $\sigma = 6$ \kms, used in much previous work and consistent
with values derived from a simple second-moment analysis, and a value
of 4 \kms, as derived above for the cool neutral ISM component in NGC
6822.

Additionally we show the \citet{joop} critical density $\log(N_H) =
20.75$, also corrected for He and metals. We compare these with the
radial distributions of H$\alpha$ and the number density of blue
stars. The radial profiles were derived using the same tilted ring
parameters that were used to derive the rotation curve.

There a few things to note: the distribution of H$\alpha$ is fairly
constant in the inner part of NGC 6822, then drops suddenly at $R
\simeq 2.0$ kpc, and runs out at $R \simeq 2.7$ kpc.  The NW companion
is visible as a very slight rise around $\sim 4.4$ kpc.  The
distribution of the number density of blue stars changes more
gradually and only reaches zero level at the edge of the \HI
disk. Note that the shape of the blue stars profile is very similar to
that of the \HI profile. This will be discussed briefly below.
Qualitatively the difference between the shape of the H$\alpha$
profile and that of the blue stellar surface density profile bears a
striking resemblance with the discrepancy between the H$\alpha$ and
GALEX UV profiles in M83 \citep{thilker05}, suggesting that the
surface density of blue stars is a good proxy for the UV surface
brightness in NGC 6822. Future UV observations will have to confirm
this.

Both the Toomre-$Q$ and the shear interpretations of the critical
density yield very similar profiles.  The exact value of the critical
density in the shear interpretation is slightly uncertain though, due
to different definitions used by different authors.  The definition
used in \citet{heb98} results in values that are $\sim 20$ per cent
lower than those derived using the \citet{joop} definition. Here we
plot only the latter.

The interpretation of these profiles depends on the value of the
velocity dispersion used. If we assume $\sigma = 6$ \kms, then we
would conclude that the disk is stable against star formation
everywhere. This would make NGC 6822 comparable with the LSB galaxies
studied by \citet{vanderhulst93}. A physical interpretation of this
result would be that just as in the LSB galaxies, local density
enhancements should provide the trigger for star formation. These
enhancements would be averaged out in a radial profile, giving the
impression of an entirely stable disk.

Assuming $\sigma = 4$ \kms, leads to a lower critical density.  For
radii $R<2$ kpc both the \citet{kennicutt89} and \citet{heb98}
critical densities follow the gas density rather closely. In this case
a possible physical interpretation would be that star formation is
self-regulating, and strives for a $Q\sim 1$ situation.  At $R \ga
2.5$ kpc the gas profile rapidly falls away from the critical density
profiles. Clearly, the outer disk is strongly sub-critical for both
values of $\sigma$, at least in the azimuthally averaged formulation
of the star formation threshold.

Note that for the 4 \kms\ case, the gas profile starts to drop away
from the Toomre-$Q$ critical density at the same radii as where the
H$\alpha$ runs out (between 2.0 and 2.2 kpc).
This is very close to the radius where the gas surface density curve
crosses the \citet{joop} critical surface density value.  For $\sigma
= 4$ \kms, both the \citet{kennicutt89} and \citet{joop} indicators of
the threshold thus give identical results, at least where the critical
radius is concerned.  Note that over the radial range examined here
the variation in gas surface density (a factor $\sim 20$) is much
larger than the variation in critical density (a factor $\sim 2.5$).
This implies that variations in gas surface density are potentially
more important in determining $\Sigma_g/\Sigma_c$ than variations in
the critical density itself.  In the following we will restrict
ourselves to the \citet{kennicutt89} threshold.

Lastly, a word of caution: as was stressed by \citet{joop}, his model
implies its own failure: when stars form, complex feedback effects and
the presence of a many-phase ISM, invalidate many of the basic
assumptions behind it.  The star formation threshold in all its
incarnations should only be regarded as a tool that indicates where
there is potential for star formation on intermediate and large scales
($\ga 0.5$ kpc or so).  Precise predictions on where the next star
forming region will form are beyond its scope.

\subsection{Blue stars and HI surface density}

In the previous sub-section we commented on the remarkable similarity
between the shapes of the radial \HI profile and the radial blue
stellar surface density profile. Here we emphasise that similarity in
a brief digression from our main star formation threshold topic.

Figure \ref{fig:hiblue} shows the inclination-corrected and helium-corrected gas
surface density profile, as well as the inclination-corrected radial
profile of the blue stellar surface density. It is clear that the
stellar surface density decreases much more rapidly than the gas
surface density profile, perhaps indicating a decreased star formation
efficiency at lower gas densities.

A simple linear scaling shows however that the shapes of these curves
are actually virtually identical.  Applying a scaling of the form
$\log(\sigma_{\rm stars}^{\rm scaled}) = 0.49 \log(\sigma_{\rm stars})
+ 0.77$ to the stellar surface density curve yields a very good match
to the gas surface density profile, with a RMS scatter of only 0.067
in $\log \sigma$, equivalent to 15 per cent.  This scaling is equivent
to a Schmidt-type law of the form $\sigma_{\rm stars} = 0.027 \cdot
\sigma_{\rm gas}^{2.04}$.  It is remarkable that the scatter in this
relation is so low and holds over 3 orders of magnitude in stellar
surface density, given the many intermediate steps needed to go from an \HI
disk to a young star (i.e.\ molecular cloud formation, collapse,
etc). Alternatively, this well-defined similarity between stellar surface
density and \HI surface density might indicate that \HI and star formation are
more intimately related than usually assumed (see
e.g. \citealt{allen}).

A similar but less well-defined relation holds between the surface
\emph{brightness} of the young stellar population and the \HI surface
density (Fig.~\ref{fig:hiblue}). The scaling does however only work
for $R < 3.2$ kpc; the blue stars in the NW companion have a surface
brightness that is over a magnitude brighter than one would derive on
the basis of the scaling.  Applying the scaling $\log(\sigma_{\rm
  stars}^{\rm scaled}) = -0.19 \cdot \mu_{\rm stars} + 5.1$, we find
that for radii $R<3.2$ kpc the RMS scatter is 0.089 in $\log \sigma$
($\sim 22$ per cent).  The relation between the total surface
brightness and the \HI surface density shows the expected declines in
densities towards the outer parts, but does not show the tight
coupling described above.

It is remarkable that in NGC 6822 we can so clearly distinguish these
various aspects of star formation: the inner parts show a well-defined
Schmidt-type law, where the surface density of young stars is directly
related to the \HI surface density, with the outer parts showing a
pronounced star formation threshold.  The latter is particularly
well-defined when one considers the local densities of gas and
stars. This is what we will return to now.

\subsection{Two-dimensional thresholds}

\begin{figure*}[t!]
\plotone{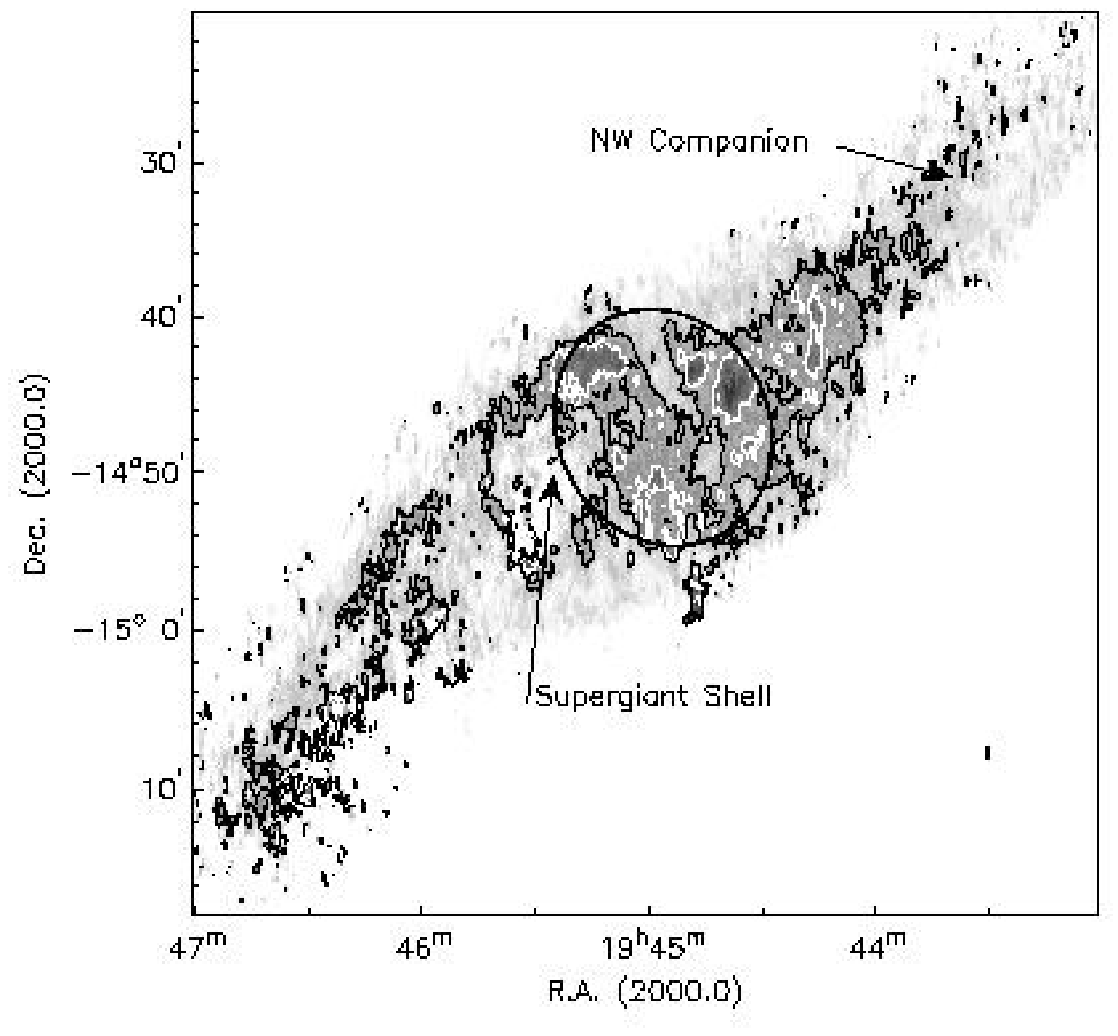} \figcaption{Second moment map derived from the \HI
  data cube. Grayscales run from 5 \kms\ (white) to 15
  \kms\ (black). The maximum values occuring in this map is 18.3
  \kms. The black contour indicates 8 \kms. The white contour
  indicates 10 \kms. Here the entire data set has been used, and no
  attempt has been made to correct for e.g.\ the presence of
  high-velocity gas or asymmetric profiles. The beam of $42.4'' \times
  12''$ is indicated in the lower-right corner. For orientation, the
  large central ellipse indicates the approximate $R_{25}$ extent as
  given in the RC3. This defines the conventional optical extent of
  NGC 6822.
\label{fig:mom2}}
\end{figure*}

\begin{figure*}[t!]
\plotone{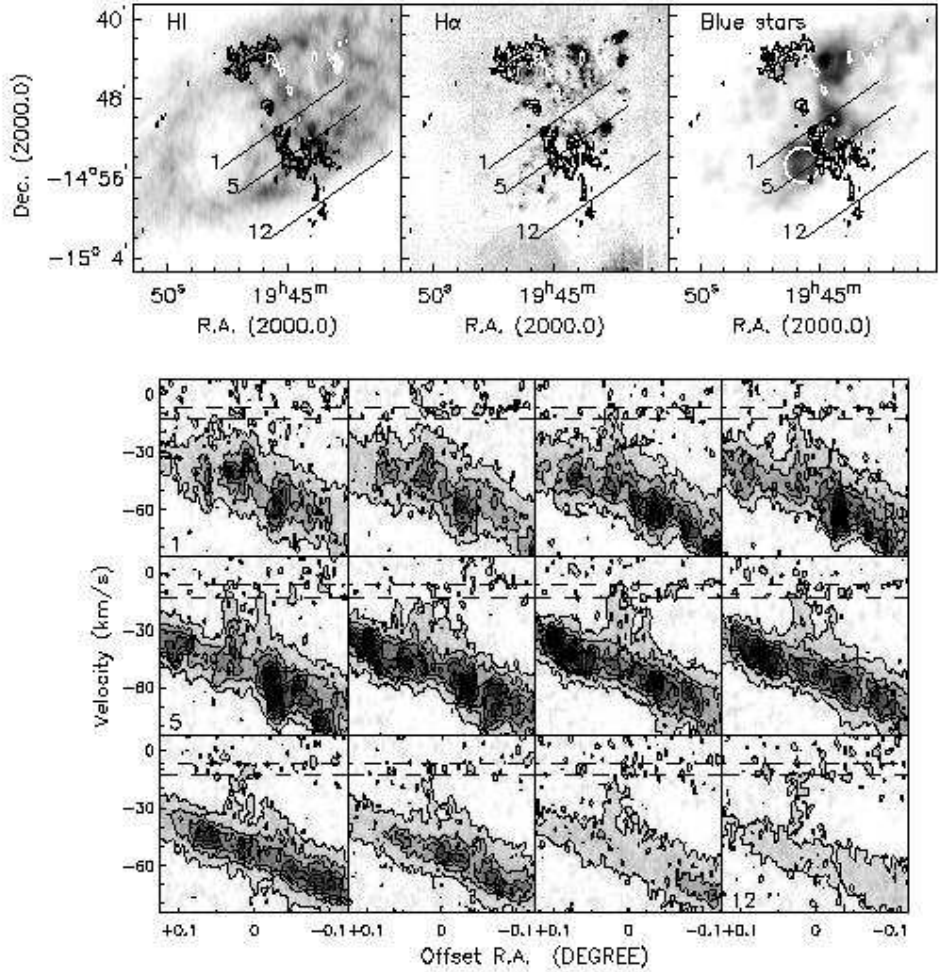} \figcaption{High-velocity gas in NGC 6822. The top
  panels show the location of all gas deviating more than $18$
  \kms\ from the rotation velocity. Black contours indicate positive
  deviation, white contours negative deviation. The top-left panel
  shows the HV gas superimposed on the total \HI map. The top-centre
  panel shows the HV gas superimposed on the H$\alpha$ map. The
  top-right shows the HVC component superimposed on the surface
  density of blue stars. The location of the Blue Plume (see text) has
  been indicated with the white circle. The bottom mosaic shows
  position-velocity slices along the lines indicated in the top
  panels. The top-left slice corresponds with line 1, the bottom-right
  one with line 12. The intermediate slices are equally spaced between
  these two limits. Slices are integrated over 12$''$ and have a
  position angle of $125^{\circ}$. The horizontal dashed lines in the
  position-velocity diagrams indicate velocities with possible
  Galactic contamination. The lowest contour level is $2\sigma$, and
  contours increase in steps of
  $4\sigma$. \label{fig:HVC}} \end{figure*}

\begin{figure*}[t!]
\plotone{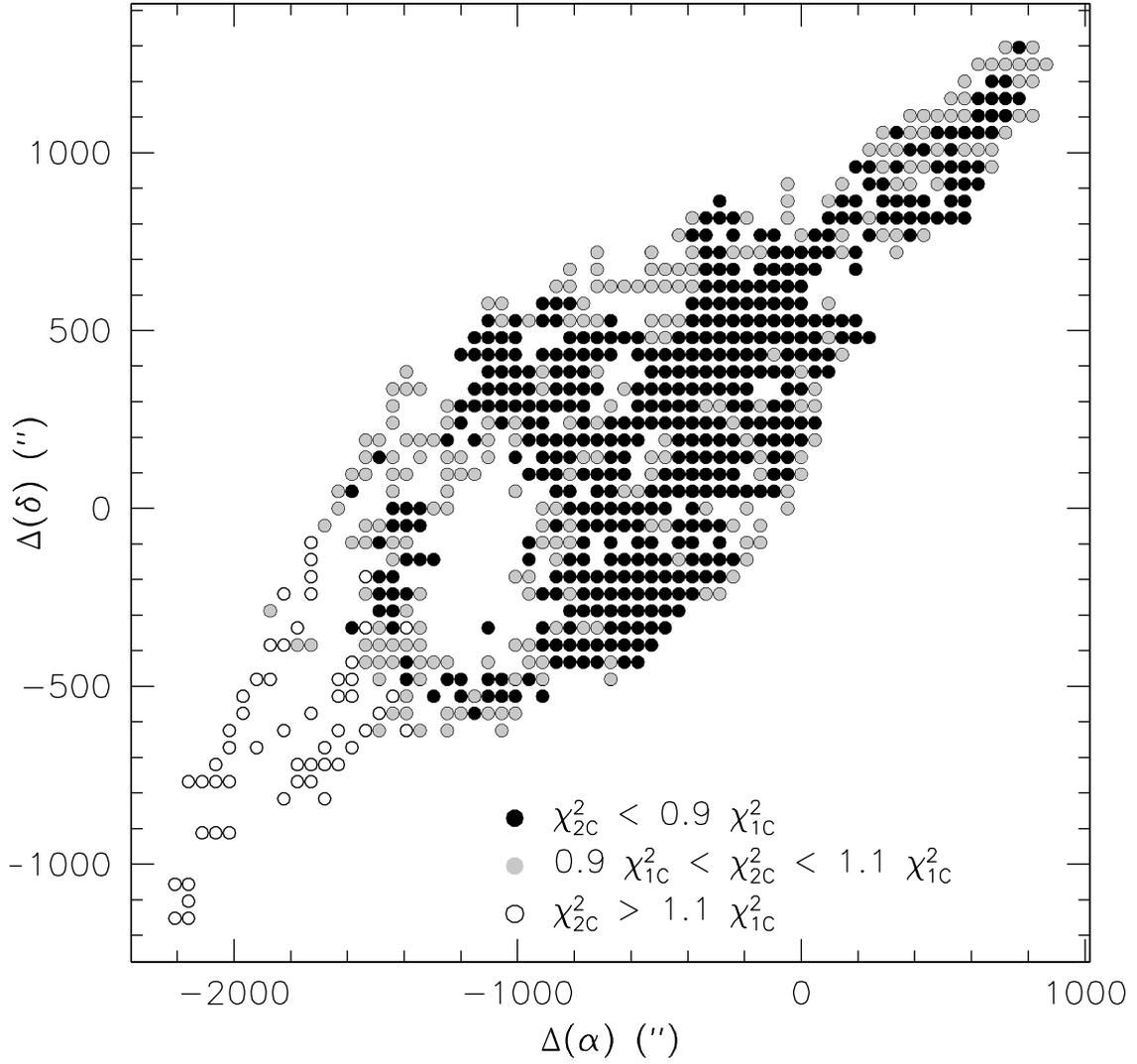} \figcaption{Comparison of the quality of one- and
  two-component Gaussian fits to the \HI velocity profiles. Filled
  black dots indicate the positions where a two-component fit is
  preferred. The open circles indicate profiles where a one-component
  fit is preferred. The grey circles show the positions where one- and
  two-component fits perform equally well. The preference for
  one-component fits in the tidal arm is an artefact of residual
  galactic emission affecting the two-component fits. Offsets are with
  respect to position $(\alpha,\delta) = (19^h44^m09^s,
  -14^{\circ}50^m54.0^s$).
\label{fig:chi2comp}} \end{figure*}

\begin{figure*}[t!]
\plotone{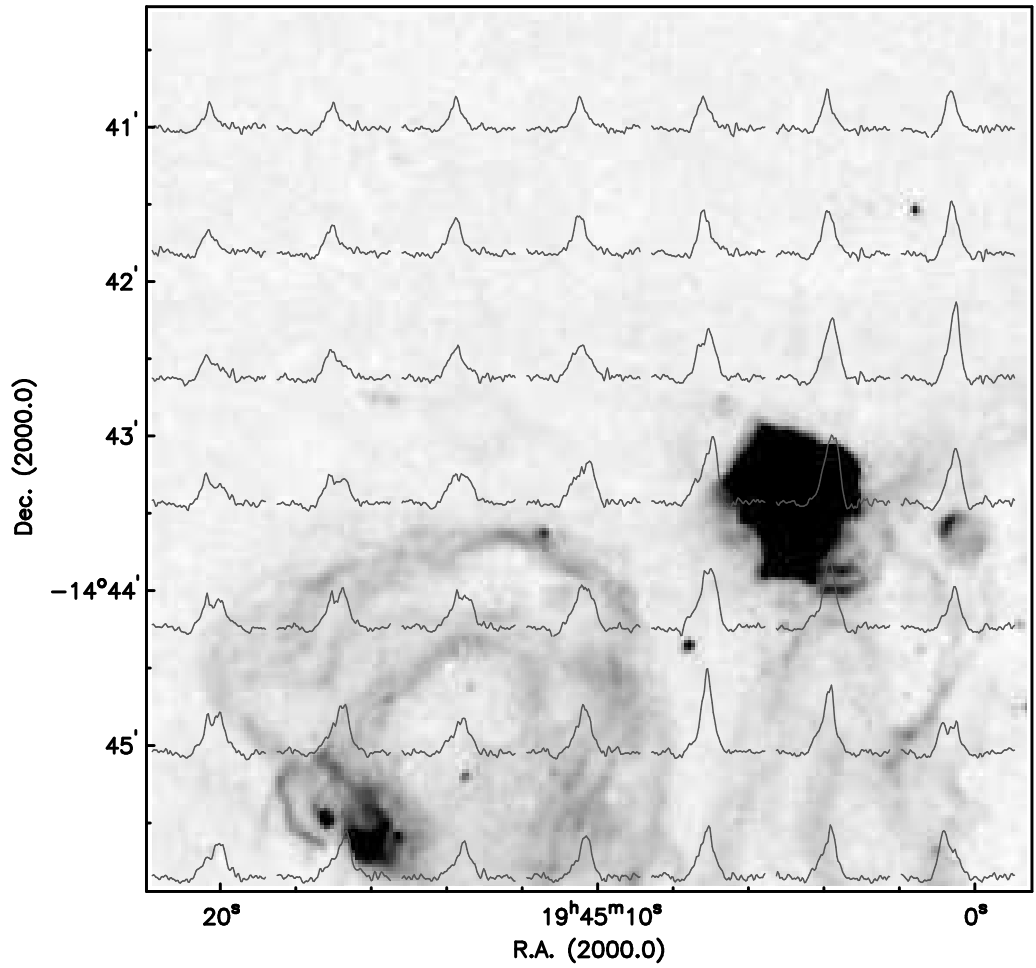} \figcaption{Examples of \HI profiles. Their position
  in the diagram corresponds with their position on the sky.  For
  comparison the corresponding local H$\alpha$ emission is shown as a
  grayscale background image. Profiles are plotted from -126 (left) to
  +35 (right) \kms.
\label{fig:prof_panel}} \end{figure*}

\begin{figure*}[t!]
\plotone{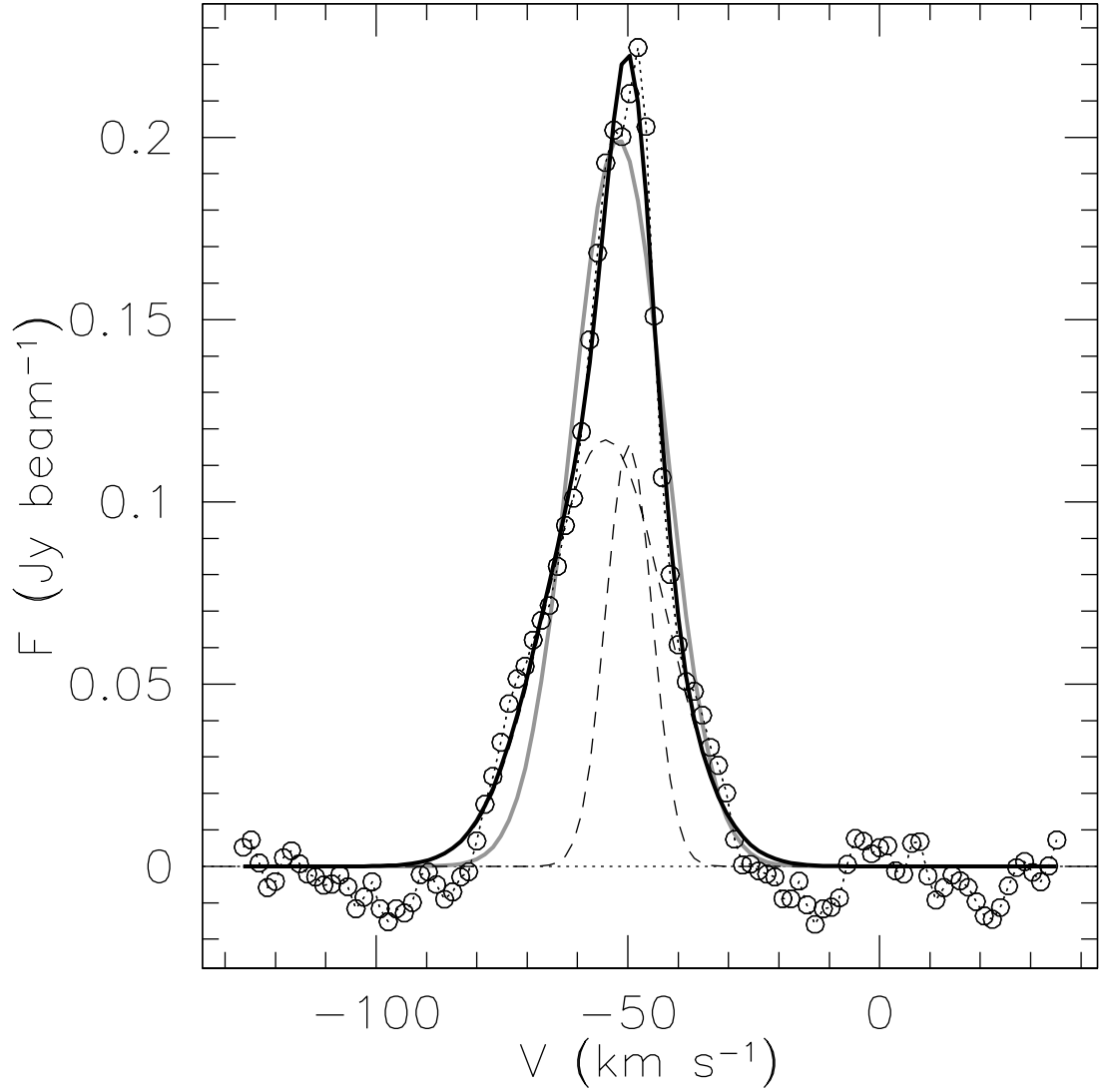}
\figcaption{Example of a non-Gaussian profile, at position ($\alpha$,
  $\delta$) = (19$^h$46$^m$05.25$^s$,
  $-14^{\circ}$45$'$10.4$''$). Open circles represent the data, the
  grey thick line a one-component Gaussian fit, the black thick line a
  two-component Gaussiaun fit. The individual components are shown are
  thin dashed curves.  The broad component has a dispersion of 11.9
  \kms, the narrow component a dispersion of 4.6 \kms.
\label{fig:prof_example}} \end{figure*}

\begin{figure*}[t!]
\plotone{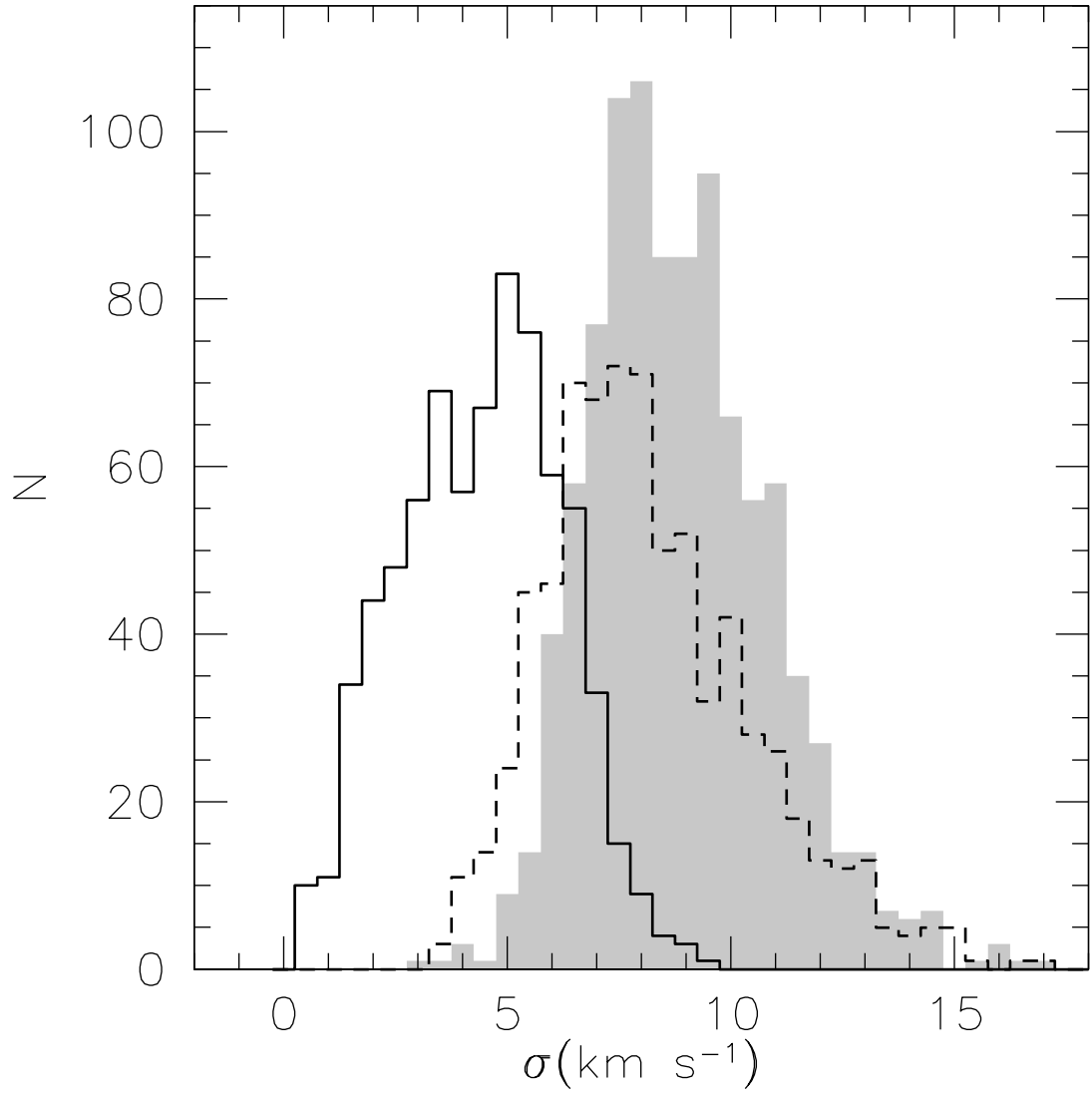}
\figcaption{Histogram of the dispersions derived from the one-component
Gaussian fits (grey histogram) and from the two-component fit, with
the broad component shown as the dashed-line histogram and the narrow
component as the full-line histogram.
\label{fig:prof_dispersion}} \end{figure*}

\begin{figure*}[t!]
\plotone{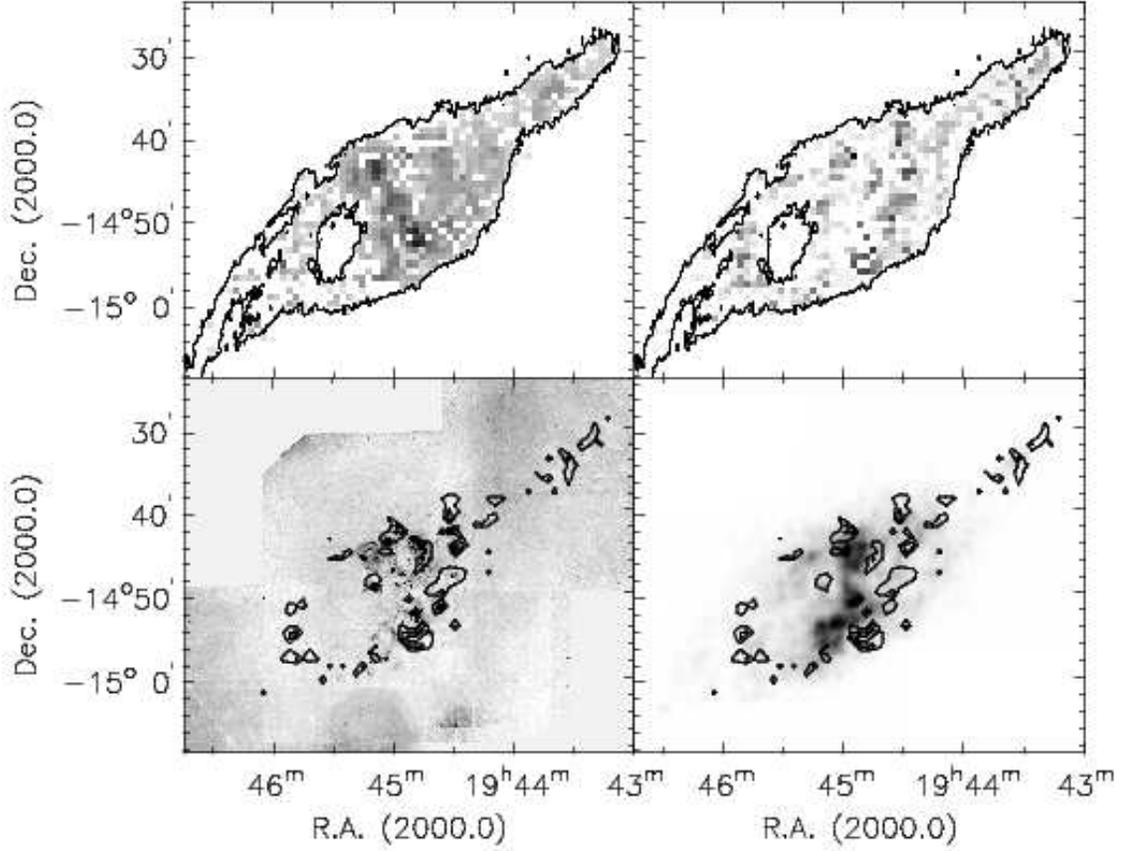} \figcaption{Distribution of the cool and warm neutral
  components of the ISM. The top-left panel shows the distribution of
  the warm, broad-dispersion component, with the grayscale running
  from 0 to $3.2 \cdot 10^{21}$ cm$^{-2}$. The contour shows the
  $5\cdot 10^{20}$ cm$^{-2}$ level of the \emph{total} \HI column
  density distribution. The top-right panel shows the distribution of
  the cool narrow-linewidth component. The grayscale runs from 0 to
  $1.7 \cdot 10^{21}$ cm$^{-2}$. The contour again shows the $5\cdot
  10^{20}$ cm$^{-2}$ level of the \emph{total} \HI column density
  distribution.  The bottom left panel compares the distribution of
  the cool component (contours) with that of the H$\alpha$
  (grayscale). The bottom-right panel shows the distribution of the
  cool component as contours overplotted on the surface density of the
  blue stars. For clarity we only show the peaks in the cool-component
  column density. In the bottom panels the contours start at $4 \cdot
  10^{20}$ cm$^{-2}$ and increase in steps of $5\cdot 10^{20}$
  cm$^{-2}$.
\label{fig:warmcool}} \end{figure*}

\begin{figure*}[t!]
\plotone{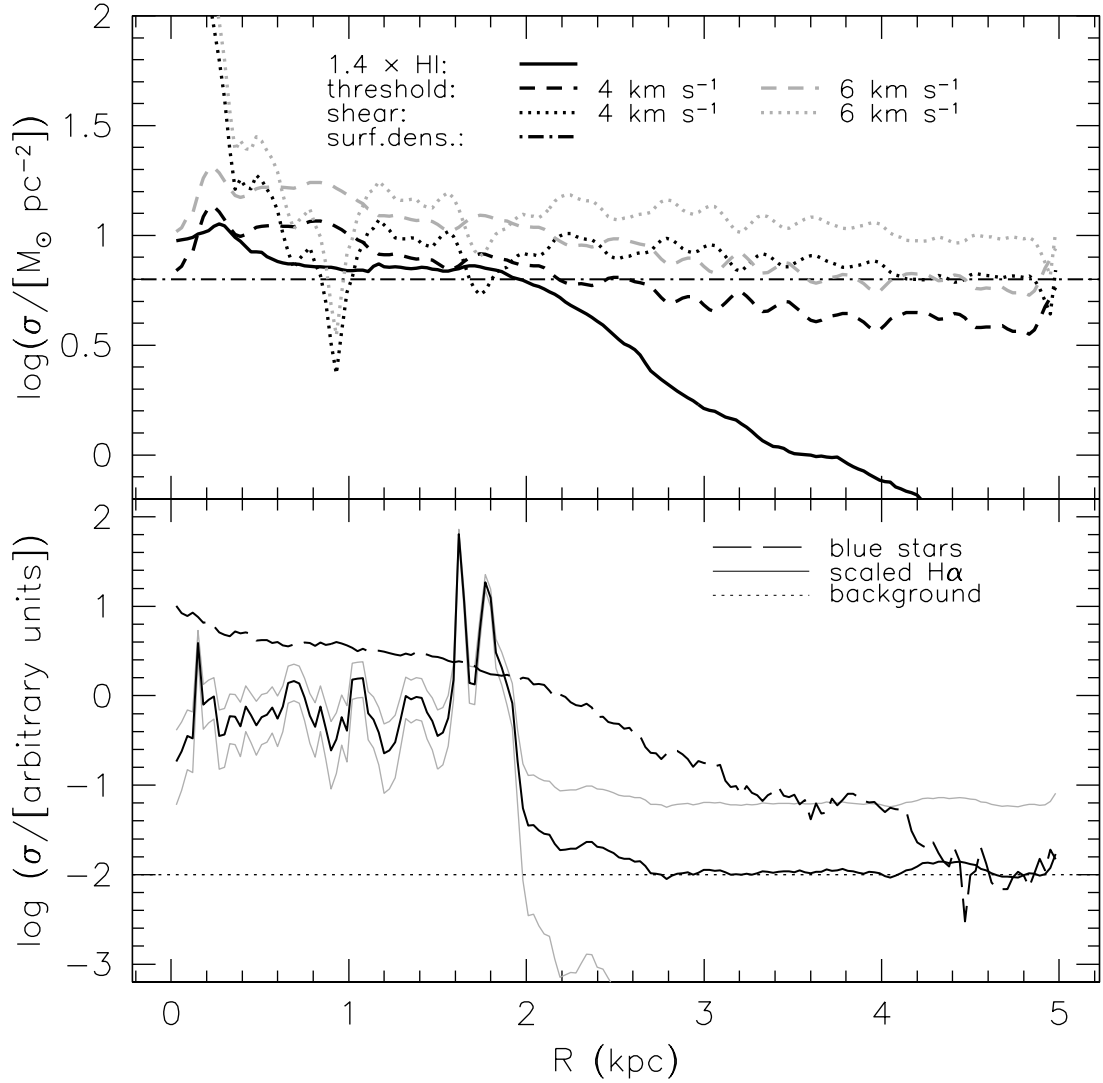} \figcaption{Comparison of the radial distributions of
  star-formation threshold related properties. In the top panel, the
  full line shows the gas surface density (corrected for helium and
  inclination). The dotted lines shows the critical density derived
  using the \citet{joop} version of the \citet{heb98} critical
  density. The dashed lines show the \citet{kennicutt89} critical
  densities. The grey critical density curves assume $\sigma = 6$
  \kms, the black critical density curves assume $\sigma = 4$ \kms.
  The dash-dotted line indicates the \citet{joop} critical density
  $\log (N_H)=20.75$ (corrected for helium). The bottom panel shows
  the radial distribution of H$\alpha$ (with the 1$\sigma$ uncertainty
  due to noise in the sky background indicated by the gray lines) and
  blue star density. The normalized background levels are also
  indicated.
  \label{fig:thresh_rad}} \end{figure*}

\begin{figure*}[t!]
\plottwo{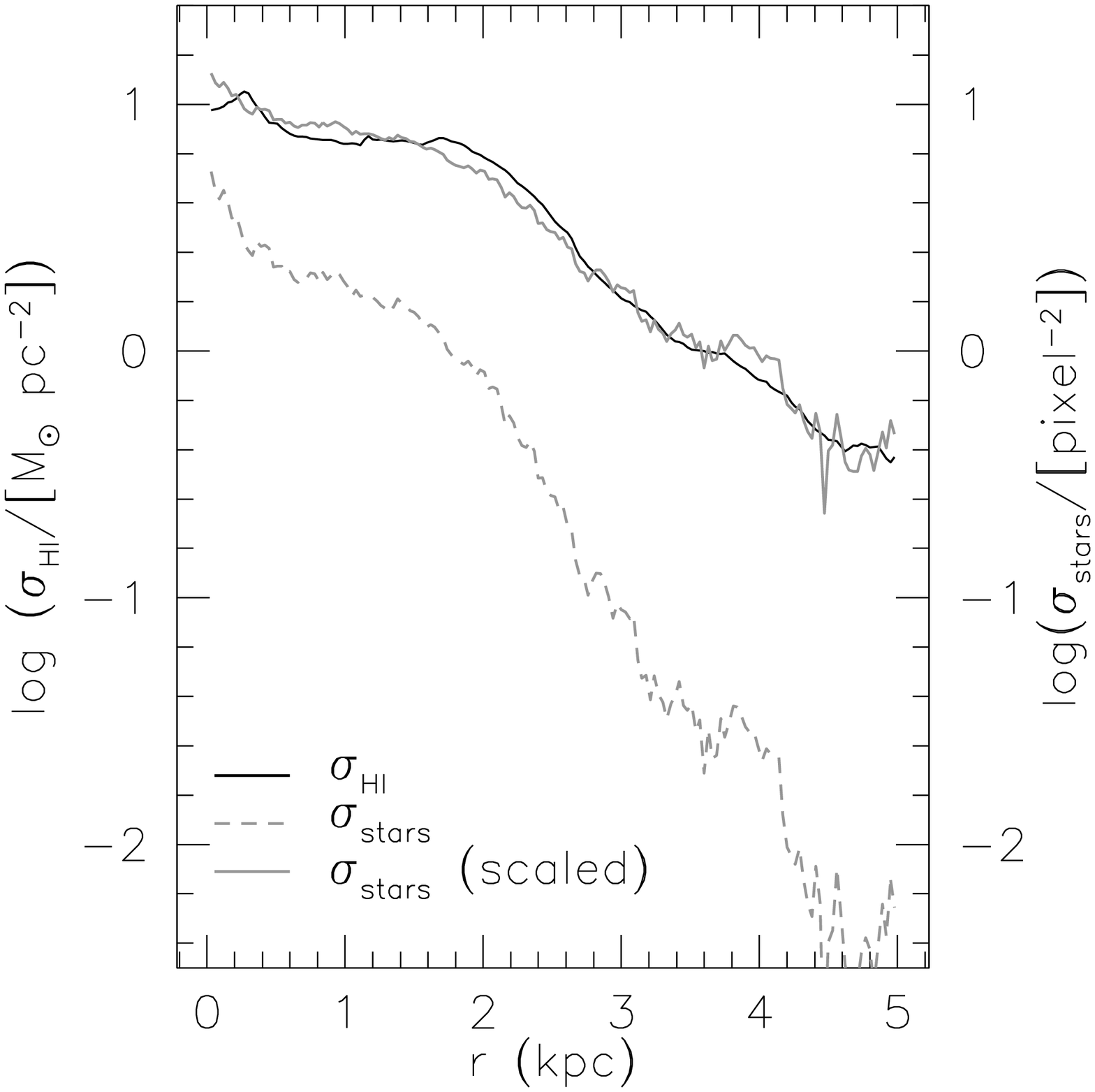}{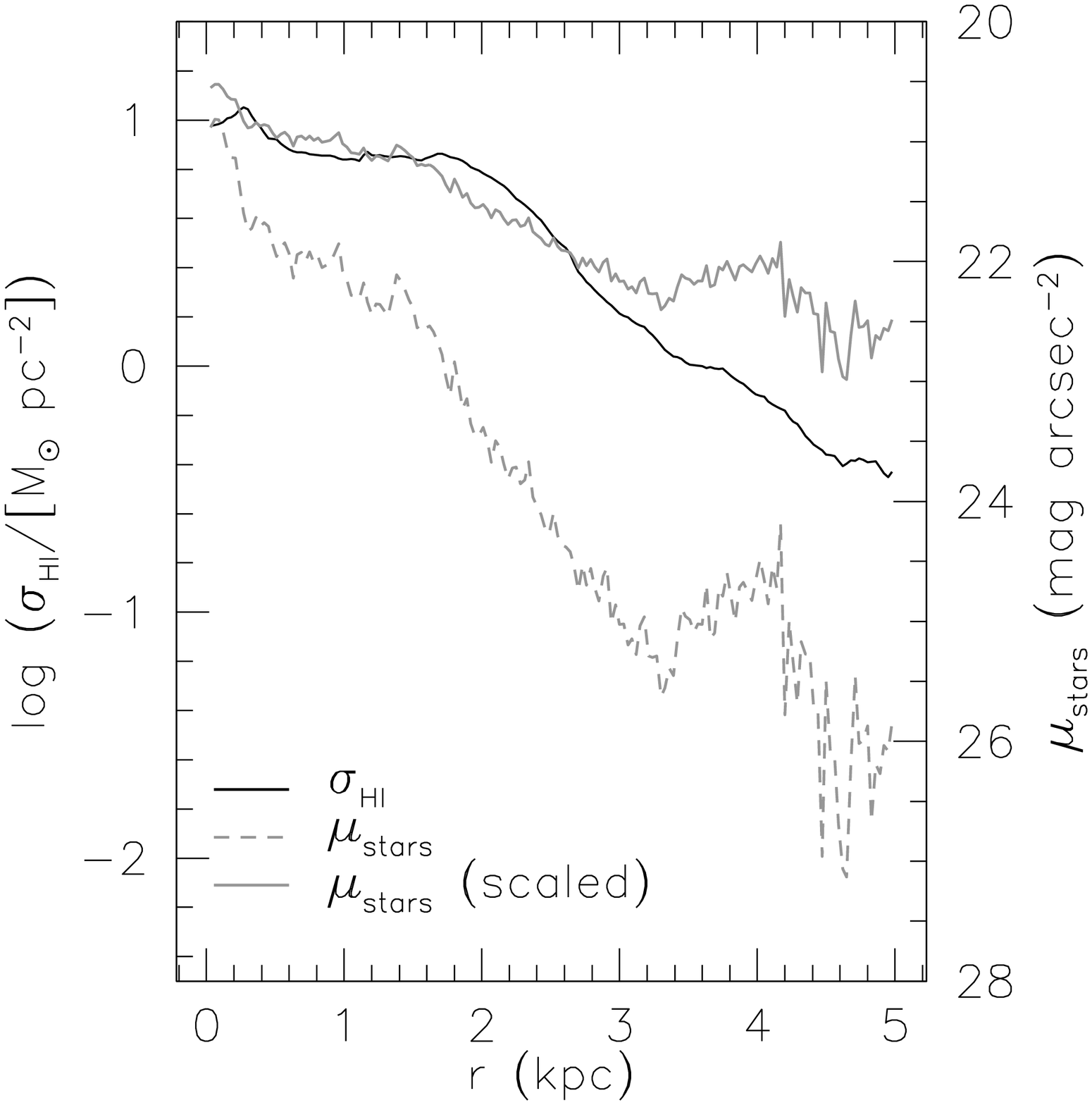} \figcaption{Comparison of the surface
  densities of gas and blue stars.  The left panel shows the
  inclination-corrected gas surface density (also corrected for
  helium) as the black line.  The dashed grey curve shows the
  logarithmic radial surface density of blue stars, also corrected for
  inclination. The full grey curve shows the scaled stellar surface
  density using a scaling $\log(\sigma_{\rm stars}^{\rm scaled}) =
  0.49 \log(\sigma_{\rm stars}) + 0.77$. The RMS scatter of the scaled
  stellar curve with respect to the gas curve is 0.067 in $\log
  \sigma$. The right panel shows a similar relation between the gas
  surface density and the inclination-corrected surface brightness of
  the blue stars. The black curve again shows the logarithmic gas
  surface density. The dashed curve shows the radial surface
  brightness profile of the blue stars (with the scale along the
  righthand axis). The grey full curve shows the scaled surface
  brightness curve $\log(\sigma_{\rm stars}^{\rm scaled}) = -0.19
  \cdot \mu_{\rm stars} + 5.1$. Here only radii $R<3$ kpc were used
  for the scaling.\label{fig:hiblue}}
\end{figure*}

With some assumptions we can investigate the local conditions for star
formation, i.e., we can extend the one-dimensional star formation
threshold to a two-dimensional equivalent.  As Eq.~(1) shows, the
relevant parameters are $\kappa$ and $c_s$ (or $\sigma$, remarks below Eq.~2
in Sect.~\ref{SEC2}).  For the dispersion we can either assume a
constant value of 4 or 6 \kms, or use the second-moment map shown in
Fig.~\ref{fig:mom2} (though as discussed in the previous sections, the
last two choices may not be as appropriate; here we show them only as
a comparison with the 4 \kms\ results).

To determine $\kappa$ we would ideally use the velocity field to
measure $V$ as a function of radius
\emph{and} position angle, i.e.\ $V= V(R,\phi)$, and so determine a 
local $\kappa$ to calculate a local critical density. Unfortunately,
due to the well-known projection effects, we can in practice only do
this with sufficient accuracy in a relatively small angular range
around the major axis. The rotational signal at other position angles
becomes too small, and even vanishes near the minor axis.

To determine a local threshold we thus have to make
the assumption that $\kappa(R,\phi) = \kappa(R)$, and use the
azimuthally averaged tilted ring rotation curve.  This potentially
limiting assumption is however more than compensated for by the fact
that we can use the two-dimensional \HI distribution to compute
$\Sigma_g/\Sigma_c$. Furthermore, the fact that an azimuthally
averaged rotation curve could be constructed with fairly small
errorbars (see \citealt{weldrake03}) already indicates that azimuthal
variations are not large, and certainly not large enough to dominate
over azimuthal variations in \HI surface density.

\begin{figure*}[t!]
\plotone{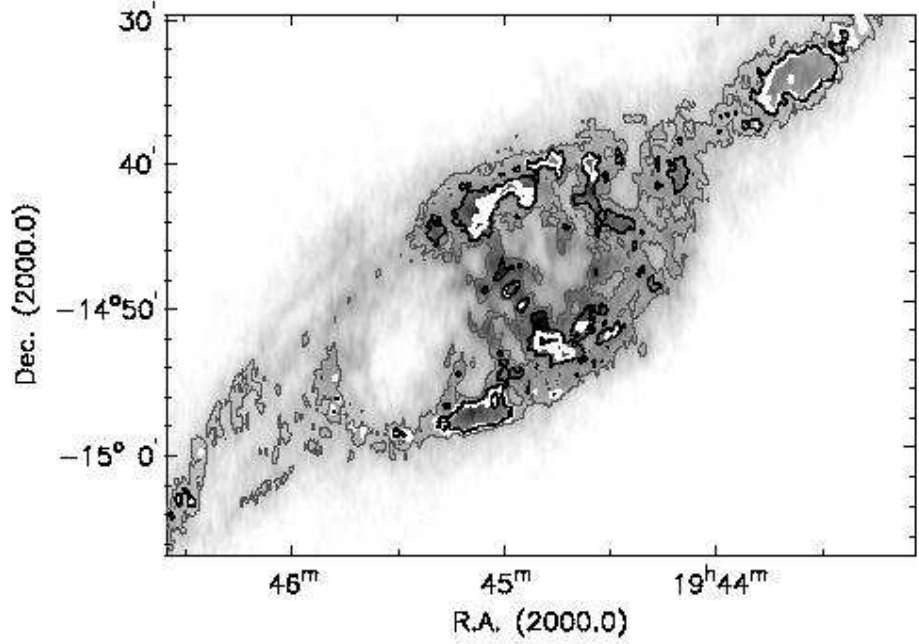} \figcaption{Comparison of the two-dimensional
  \citet{kennicutt89} star formation thresholds.  The grayscale image
  shows the total \HI map. The contours enclose the regions predicted
  to be unstable to star formation.  Various assumptions for the
  velocity dispersion are shown. The white contours have been computed
  using the variable dispersion as shown in the second-moment map. The
  black contours enclose the unstable regions computed assuming a
  constant dispersion of 6 \kms. The grey contours enclose the
  unstable regions computed using a dispersion of 4 \kms.
 \label{fig:thresh_q}} \end{figure*}

We have computed maps of the local ratios of
$\Sigma_g/\Sigma_c$. Figure \ref{fig:thresh_q} shows
$\Sigma_g/\Sigma_c = 1$ contours overlaid on the \HI surface density
map for the cases of constant velocity dispersions $\sigma = 4$ and 6
km s$^{-1}$, as well as a variable dispersion derived from
the second-moment map in Fig.~\ref{fig:mom2}.

The conclusions are similar to the azimuthally averaged case discussed
above. For the $\sigma = 6$ \kms\ case, only a small fraction of the
disk (mostly towards the edge of the \HI disk) is unstable to star
formation. The variable dispersion case does not improve matters, the
higher dispersions in the inner part of the galaxy make the gas there
apparently more stable against collapse.

In the constant dispersion cases, and most prominently for the 4
\kms\ case, a much larger fraction of the disk now has the potential
to participate in star formation.  As the surface density criterion is
a local criterion it can also be applied to the NW companion, even
though it is an independent system.  We see that the NW companion is
indeed unstable to star formation, as confirmed by the presence of
H$\alpha$ and young stars there.

\begin{figure*}[t!]
\plotone{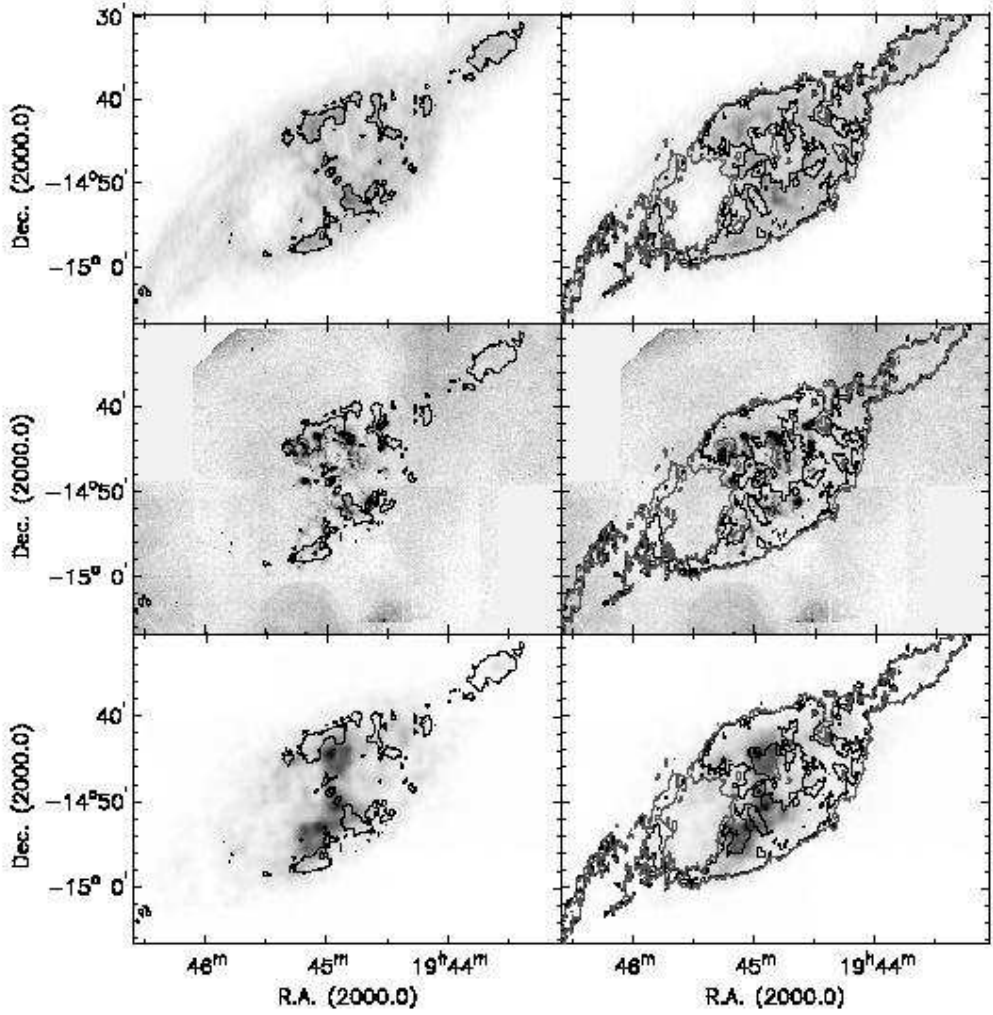}
\figcaption{Comparison of the various local star formation threshold
  predictions. The left column shows predictions for a $\sigma = 6$
  \kms\ Toomre-$Q$ threshold. The black contour encloses unstable
  $\Sigma_g/\Sigma_c \ge 1$ regions. The black contour in the right
  column shows the prediction for a Toomre-$Q$ threshold assuming
  $\sigma = 4$ \kms. The grey contour in the right column shows the
  critical \citet{joop} surface density $\log(N_{HI}) = 20.75$.  The
  top row compares the predictions with the total \HI distribution
  (grayscales). The centre panels compare with the H$\alpha$
  distribution, the bottom panels with the blue star surface density
  distribution (both in grayscales). Note the good correspondence
  between the surface density and the Toomre-$Q$ predictions in the
  right-hand column.
  \label{fig:thresh_schaye}} \end{figure*}

In the following we will restrict ourselves to the constant dispersion
thresholds.  In Fig.~\ref{fig:thresh_schaye} we compare the
distribution of $\Sigma_g/\Sigma_c$ with those of H$\alpha$ and the
blue stars as well. The left panels show $\Sigma_g/\Sigma_c = 1$
contours for a 6 \kms\ dispersion. The right-hand panels show the same
for a 4 \kms\ dispersion. Overall, the 4 \kms\ contours give a much
better description of the locations of (recent) star formation
activity than the 6 \kms\ contours.  Note the anticorrelation between
blue star density and unstable regions for the 6 \kms\ case.  Overall,
the threshold seems to work best in the outermost parts.  Formally
speaking the epicycle frequency diverges at small $R$ and the
assumption that the disk is thin breaks down.  \citet{deblokwalter00}
derive a scale height of 0.28 kpc for NGC 6822, meaning that for $R
\la 0.5$ kpc the threshold analysis is not valid.

The right-hand panels of Fig.~\ref{fig:thresh_schaye} also show the
surface density threshold described in \citet{joop}. Plotted is the
$\log (N_{HI}) = 20.75$ surface density contour (corrected for
inclination and assuming that \HI is the main component of the H
distribution).  The unstable area comfortably encompasses the
H$\alpha$ and blue stars. Note the very good correspondence with the 4
\kms\ dispersion contours (although the surface density criterion
seems to perform slightly better in the inner parts compared with the
distribution of blue stars).  Just as in the azimuthally averaged
picture, we again see that the critical radii in both cases become
equivalent when the proper choice for velocity dispersion is
made. Also note that the critical radius as defined here corresponds
very closely with the edge of the \HI disk.

We can investigate whether these conclusions also hold locally by
looking at the pixel-to-pixel correlations between the various
components. We have regridded the HI, H$\alpha$ as well as the
critical density and blue star density images to a common pixel size
of 12$''$.

\begin{figure*}[t!]
\plotone{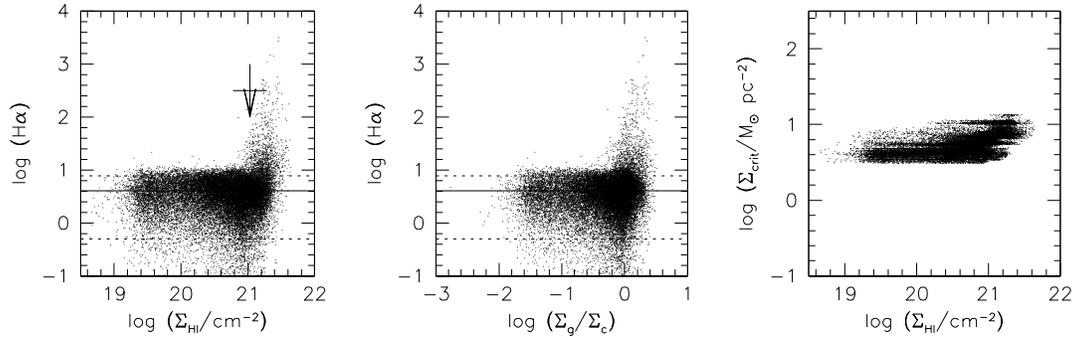}
\figcaption{Pixel-to-pixel correlations. The left  panel shows the
  correlation between H$\alpha$ and \HI surface density. The arrow
  indicates the $\log(N_{HI}) = 20.75$ level (corrected for
  inclination) with the associated uncertainty, as described in
  \citet{joop}. The full horizontal line indicates the average
  background level, the dotted lines the 1$\sigma$ noise in the
  background level. The centre panel shows the correlation between
  H$\alpha$ intensity and $\Sigma_g/\Sigma_c$.  Lines are as in the
  top-left panel. The right panel shows the correlation between
  \HI column density and critical density $\Sigma_c$.
  \label{fig:pixcor}} \end{figure*}

The left panel in Fig.~\ref{fig:pixcor} shows the local correlation
between \HI and H$\alpha$ on scales of $12''$ ($\sim$ 30 pc).  The
1$\sigma$ noise in the background sky level is indicated in the
figure.  Again we see clear evidence of a well-defined star formation
threshold. At $\log(N_{HI}) \approx 20.9$ (20.7 after inclination
correction) the maximum H$\alpha$ flux per pixel suddenly increases
sharply (roughly 1 dex over only 0.15 dex increase in \HI column
density).  At lower densities only very marginal amounts of H$\alpha$
are present. The arrow in the left panel indicates the predicted value
of the \citet{joop} threshold with the associated errorbars
indicated. There is again good agreement between observed and
predicted values.

The centre panel shows the pixel-to-pixel relation between
$\Sigma_g/\Sigma_c$ and H$\alpha$, assuming $\sigma = 4$ \kms.  Again,
the two threshold alternatives yield very similar results.  In the
right panel in Fig.~\ref{fig:pixcor} we show the relation between
$\Sigma_{HI}$ and $\Sigma_c$.  Both axes span the same range, but we
see that the variation in critical density is much smaller (by a
factor $\sim 200$) than that in \HI density. This behaviour was
already discussed for the azimuthally averaged case, but is now also
seen to hold locally.

These results again show that changes in $\Sigma_g/\Sigma_c$ are
completely dominated by changes in $\Sigma_g$ and that, at least in
the case of NGC 6822, the gas density seems to be the ultimate driver
for star formation, independent of rotation. Interpreting this in the
framework of the \citet{joop} models, the relevant value of $Q$ is
completely determined by $\sigma$, and its steep drop  at the
onset of the cool phase. The variation in the other parameters that
form $Q$ are not important. 

\section{Summary}

We have presented a comprehensive analysis of the star formation
threshold in the Local Group dwarf galaxy NGC\,6822. The absence of
complicating spiral density waves, bar motions, strong shear, etc.,
make NGC\,6822 an ideal target for such a study.  The proximity of the
galaxy furthermore allows us to investigate the threshold at very high
spatial resolution ($<100$ pc).

Regions with High Velocity gas (i.e., with velocities deviating more
than 18 km\,s$^{-1}$ from the rotation velocity) were found to be in
the vicinity of current star formation (as traced through H$\alpha$)
although no one-to-one correspondence exists. This is exemplified by
the presence of an \HI `chimney' for which no single parental
cluster/association can be easily identified, although it is likely
that the combined effects of a larger-scale region of recent
star-formation may have created this feature.

The cool and warm phases of the \HI are identified by fitting
two-component Gaussians to the \HI profiles. In doing so we find two
components, one broad component with a mean dispersion of
8\,km\,s$^{-1}$ and a narrow component at 4\,km\,s$^{-1}$. A
comparison with the H$\alpha$ maps shows that every region with a
dominant low-dispersion \HI peak has a compact H{\sc ii} region in its
immediate environment.

We have compared various descriptions of the star formation threshold
(Toomre-$Q$, shear, constant density threshold) to radial averages of
the \HI and ongoing star formation (as traced by H$\alpha$) and find
that of the thresholds that are based on the dynamics of the galaxy,
the Toomre criterion with a low velocity dispersion of 4\,km\,s$^{-1}$
(our `cool' component) fits the distribution of star formation in
NGC\,6822 best.  However, the predictions of that model are virtually
identical to those of the model assuming a simple constant star
formation surface density threshold of $\log N_{HI}=20.75$.

Part of the agreement is due to the fact that the critical density
$\Sigma_{\rm crit}$ as derived through, e.g., the Toomre criterion,
has a much smaller dynamic range than the measured gas surface
densities $\Sigma_{\rm gas}$.  Changes in the ratio $\Sigma_{\rm
  crit}/\Sigma_{\rm gas}$ are thus completely dominated by the change
in $\Sigma_{\rm gas}$.  For the radial profiles, we also find a
striking correspondence between the \HI density and the number density
of blue stars, showing that these two components are intimately linked
to each other in NGC\,6822.

Our data also allow us, for the first time, to calculate
high-resolution two-dimensional maps of the critical density in
NGC\,6822. Our finding is similar to those for the radial profiles, in
the sense that a Toomre criterion (using a dispersion of
4\,km\,s$^{-1}$) gives very similar results compared to a simple star
formation threshold of $\log(N_{HI})=20.75$, also locally. This
emphasizes that the star formation threshold in NGC 6822 does not
depend on rotation but is a purely local phenomenon.  One needs to
keep in mind though that the current analysis was done for one dwarf
galaxy only. Even though the observations are exquisite in the details
they show, results drawn from this study will need to be confirmed by
similar analyses on other dwarf galaxies.

\clearpage

\clearpage 

\clearpage

\end{document}